\documentclass[12pt,a4paper]{article}
\usepackage[T1]{fontenc}
\usepackage{lmodern}
\usepackage[a4paper,hmargin=23.5mm,top=24mm,textheight=230mm]{geometry}
\usepackage{amsmath,amssymb}
\usepackage{graphicx}
\usepackage{subfigure}
\usepackage{booktabs}
\usepackage{placeins}
\usepackage{indentfirst}
\usepackage{cite}
\usepackage{microtype}
\usepackage[colorlinks=true,linkcolor=blue,urlcolor=blue,citecolor=red,
  hypertexnames=false,bookmarksopen=true,pdfstartview=FitV,pdfpagemode=UseNone,
  pdftitle={Radial stability of nonlinearly scalarized black holes in Einstein-phantom scalar-Gauss-Bonnet theory},
  pdfauthor={De-Cheng Zou, Yun Soo Myung, Meng-Yun Lai, Hyat Huang, Wen-Cong Gan}]{hyperref}

\newcommand{\ud}{\mathop{}\!\mathrm{d}}

\begin{document}

\renewcommand{\thefootnote}{\fnsymbol{footnote}}
\begin{center}
{{{\Large \bf Radial stability of nonlinearly scalarized black holes in Einstein-phantom scalar-Gauss-Bonnet theory}}}\\[8mm]

De-Cheng Zou$^{a}$\footnote{dczou@jxnu.edu.cn}, Yun Soo Myung$^{b}$\footnote{Corresponding author: ysmyung@inje.ac.kr}, Meng-Yun Lai$^{a}$\footnote{mengyunlai@jxnu.edu.cn}, \\ Hyat Huang$^{a}$\footnote{hyat@mail.bnu.edu.cn} and Wen-Cong Gan$^{a}$\footnote{ganwencong@jxnu.edu.cn}\\[7mm]
${}^{a}$School of Physics, Jiangxi Normal University, Nanchang 330022, China\\
${}^{b}$Center for Quantum Spacetime, Sogang University, Seoul 04107, Republic of Korea
\end{center}
\renewcommand{\thefootnote}{\arabic{footnote}}
\setcounter{footnote}{0}

\begin{abstract}
We study the radial stability of nonlinearly scalarized black holes found in
Einstein--phantom scalar--Gauss--Bonnet theory with polynomial and exponential
couplings. Two branches of lower and primary are found for phantom scalarized black holes.
The coupled scalar--metric perturbations reduce to a
Schrödinger-type equation, after eliminating redundant metric perturbations. The lower branch including the pure quartic
solutions has one growing mode, while positive zero-energy solutions for the primary branch
could establish stability  through the $S$-deformation
method.   Stable primary solutions are found for the exponential coupling. 
Also, we explore the connection between radial stability and thermodynamic analysis for two branches. 
Furthermore, we show that  canonical scalarized black holes obtained from Einstein--scalar--Gauss--Bonnet theory with  polynomial coupling
 reveals  unstable lower and stable primary branches.
\end{abstract}

\section{Introduction}
\label{intro}

Scalar fields appear in many contexts of gravitational physics. In Einstein
gravity minimally coupled to a scalar field, however, the no-hair theorems
rule out black holes with scalar hair
\cite{Bekenstein:1974sf,Bekenstein:1975ts,Bekenstein:1995un}.
These theorems can be circumvented in various ways, for instance when the
scalar field couples nonminimally to the curvature
\cite{Herdeiro:2015waa,Kanti:1995vq,Sotiriou:2013qea}.
Long ago, Damour and Esposito-Far\`ese found the mechanism of spontaneous
scalarization in scalar--tensor theories when studying neutron stars
\cite{Damour:1993hw}. An analogous curvature-induced mechanism arises in
Einstein--scalar--Gauss--Bonnet (EsGB) theory, where a tachyonic instability
of the Schwarzschild black hole triggers the growth of the scalar field, and
branches of scalarized black holes bifurcate from the bald solution
\cite{Doneva:2017bvd,Silva:2017uqg,Antoniou:2017acq,Antoniou:2017hxj}.
The stability of these branches has since been examined through scalar,
radial, axial and polar perturbations
\cite{Myung:2018iyq,Blazquez-Salcedo:2018jnn,Silva:2018qhn,
Blazquez-Salcedo:2020rhf,Blazquez-Salcedo:2020caw,Minamitsuji:2024twp}.
See Ref.~\cite{Doneva:2022ewd} for a review. Moreover, the effects of a scalar
mass and of self-interaction on scalarization, as well as scalar modes,
have been investigated
\cite{Doneva:2019vuh,Macedo:2019sem,Peng:2020znl,Macedo:2020tbm}.
Spontaneous scalarization is not restricted to static black holes. It also
occurs for rotating black holes, where the spin itself can induce
scalarization \cite{Cunha:2019dwb,Collodel:2019kkx,Dima:2020yac,Hod:2020jjy,
Doneva:2020kfv,Doneva:2020nbb,Herdeiro:2020wei,Berti:2020kgk},
and it has been extended to Kerr--Newman and Taub--NUT black holes
\cite{Lai:2022ppn,Liu:2024bzh}.

Nonlinear scalarization provides another route to scalarized black holes. When the
second derivative of the coupling function vanishes at $\phi=0$, the
effective mass term in the linearized scalar equation is absent, and the
Schwarzschild black hole is stable against linear scalar perturbations.
Nevertheless, Doneva and Yazadjiev showed that a sufficiently large scalar
perturbation can activate the higher-order terms of the coupling and lead to
scalarized black holes \cite{Doneva:2021tvn}. This nonlinear mechanism has
since been studied for black holes and exotic compact objects
\cite{Zhang:2023jei,Pombo:2023lxg}. For the bounded exponential coupling, a
radial mode analysis in the canonical theory revealed both stable and
unstable branches \cite{Blazquez-Salcedo:2022omw}. Nonlinear scalarization
was also found for multi-scalar Gauss--Bonnet black holes
\cite{Staykov:2022uwq,Staykov:2024jbq} and for Kerr black holes
\cite{Doneva:2022yqu}, and rotating and supermassive nonlinearly scalarized
black holes have been constructed \cite{Lai:2023gwe,Liu:2025eve}.
Canonical black holes with polynomial couplings, including the pure quartic
limit, were constructed in Ref.~\cite{Zou:2024paper}. The metric backreaction
makes their branch structure depend on the higher-order coefficient, but
their radial stability was not analyzed in that work.

Scalar fields with a negative kinetic term, called phantom fields,
were introduced as dark energy with an equation of state $w<-1$
\cite{Caldwell:1999ew}. In black hole physics, the accretion of phantom
energy decreases the black hole mass \cite{Babichev:2004yx}, and phantom
scalar fields support the Ellis--Bronnikov wormholes
\cite{Ellis:1973yv,Bronnikov:1973fh}, regular black holes
\cite{Bronnikov:2005gm} and phantom black holes in Einstein--Maxwell--dilaton
theory \cite{Clement:2009ai}. The
Ellis--Bronnikov wormholes have one growing spherical mode
\cite{Gonzalez:2008wd}, and instabilities were also found for other phantom
wormholes and for regular black holes \cite{Bronnikov:2012ch}. Phantom
scalarization with a different coupling was considered in
Ref.~\cite{Li:2025vcq}. In the phantom theory considered here, both the
scalar kinetic term and the Gauss--Bonnet term change sign. The
probe scalar equation is then the same as in the canonical theory, but
the metric backreaction is different.

In the present work, we study the radial stability of nonlinearly scalarized
black holes in Einstein--phantom scalar--Gauss--Bonnet theory. We use the
polynomial and exponential solutions constructed in
Ref.~\cite{Zou:2026staticthermo}, where their thermodynamics was also studied.
The phantom lower branch has larger Wald entropy at fixed mass and lower
free energy at fixed temperature than Schwarzschild black holes, opposite
to the canonical lower branches \cite{Zou:2026thermo,Herdeiro:2026sur}.
The coupled scalar--metric perturbations allow us to examine how this
thermodynamic ordering relates to  dynamical stability.

The paper is organized as follows. In Sec.~\ref{radialframework}, we
explain briefly  how to obtain the phantom scalarzied black hole solutions, and formulate the radial perturbation scheme.
Section~\ref{radialstability} presents the stability results for two (lower and primary)  branches and their
comparison with canonical primary  branch. Also, we discuss comparison of radial stability with thermodynamic analysis. 
We conclude our results in Sec.~\ref{con}. Appendix~\ref{radialderivation} includes  the
derivation of the radial perturbation equation.

\section{Black hole solutions and radial perturbations}
\label{radialframework}

\subsection{Phantom scalarzied  black hole solutions}
\label{model}

The four-dimensional Einstein--scalar--Gauss--Bonnet action for the canonical
and phantom theories is given by~\cite{Zou:2024paper,Zou:2026staticthermo}
\begin{eqnarray}
&&S=\frac{1}{16\pi}\int\ud^4x\sqrt{-g}\left[R+\epsilon\left(-2\nabla_\mu\phi\nabla^\mu\phi+\lambda^2\zeta(\phi){\cal R}_{\rm GB}^2\right)\right],
\label{action}
\end{eqnarray}
where $\epsilon=+1$ denotes the canonical theory and $\epsilon=-1$
the phantom theory. The Gauss--Bonnet invariant is
\begin{eqnarray}
&&{\cal R}_{\rm GB}^2=R_{\mu\nu\rho\sigma}R^{\mu\nu\rho\sigma}-4R_{\mu\nu}R^{\mu\nu}+R^2.
\label{rgb}
\end{eqnarray}

Varying the action with respect to the metric and scalar field gives
\begin{eqnarray}
&&G_{\mu\nu}+\epsilon\Gamma_{\mu\nu}
=\epsilon\left(2\nabla_\mu\phi\nabla_\nu\phi
-g_{\mu\nu}\nabla_\alpha\phi\nabla^\alpha\phi\right),
\label{Einstein}\\
&&\Box\phi+\frac{\lambda^2}{4}\zeta'(\phi){\cal R}_{\rm GB}^2=0,
\label{KG}
\end{eqnarray}
where the Gauss--Bonnet contribution to the metric equation is
\begin{align}
\Gamma_{\mu\nu}=4\lambda^2\bigg[&
 R_{\mu\rho\nu\sigma}\nabla^\rho\nabla^\sigma\zeta
 +2R_{\rho(\mu}\nabla_{\nu)}\nabla^\rho\zeta
 -g_{\mu\nu}R_{\rho\sigma}\nabla^\rho\nabla^\sigma\zeta\nonumber\\
 &-R_{\mu\nu}\Box\zeta
 +\frac R2\left(g_{\mu\nu}\Box\zeta-\nabla_\mu\nabla_\nu\zeta\right)\bigg],
\label{Gamma}
\end{align}
with $\zeta=\zeta(\phi)$, and the parentheses denote symmetrization.
The scalar equation contains no explicit $\epsilon$, but the metric
response depends on this sign.

For nonlinear scalarization, we consider the couplings
\begin{eqnarray}
&&\zeta_p(\phi)=\alpha\phi^4-\beta_p\phi^8,\qquad
\zeta_e(\phi)=\frac{1-e^{-\beta_e\phi^4}}{4\beta_e},
\label{couplings}
\end{eqnarray}
which both satisfy $\zeta''(0)=0$, so the bald Schwarzschild solution
has no tachyonic scalar mode. Their Taylor expansions agree through order
$\phi^8$ when $\alpha=1/4$ and $\beta_e=8\beta_p$.

The static black holes of Ref.~\cite{Zou:2026staticthermo} have the metric
\begin{eqnarray}
&&\ud s^2=-A(r)\ud t^2+\frac{\ud r^2}{B(r)}+r^2\ud\Omega_2^2.
\label{staticmetric}
\end{eqnarray}
Near a regular event horizon, the fields have the expansions
\begin{eqnarray}
&&A=a_1(r-r_H)+\cdots,\quad B=b_1(r-r_H)+\cdots,\quad \phi=\phi_H+\phi'_H(r-r_H)+\cdots
\label{horizonexp}
\end{eqnarray}
where the field equations fix
\begin{eqnarray}
&&b_1=\frac{1}{r_H+2\epsilon\lambda^2\zeta'(\phi_H)\phi'_H},\qquad
\phi'_H=-\frac{6\lambda^2\zeta'(\phi_H)}{r_H^3(1+\sqrt{\Delta_H})},
\nonumber\\
&&\Delta_H=1-\frac{24\epsilon\lambda^4[\zeta'(\phi_H)]^2}{r_H^4}.
\label{deltah}
\end{eqnarray}
For the phantom sign, $\Delta_H\geq1$, so the canonical horizon bound
$\Delta_H=0$ is absent.

At infinity,
\begin{eqnarray}
&&A=1-\frac{2M_A}{r}+O(r^{-2}),\qquad
B=1-\frac{2M_B}{r}+O(r^{-2}),\qquad
\phi_0=\frac{Q_s}{r}+O(r^{-2})
\label{asymptotic}
\end{eqnarray}
The time coordinate is normalized by $A(\infty)=1$, and the constant
scalar mode vanishes. The coefficients $M_A$ and $M_B$ give two estimates
of the ADM mass. We use $M\equiv M_A$, and $Q_s$ denotes the scalar charge.
We measure lengths in units of $\lambda$ and set $\lambda=1$ below.

The phantom solutions considered here have nodeless scalar profiles because they are obtained through nonlinear scalarization. When
two solutions coexist at fixed $r_H$, we call the branches with larger and
smaller horizon scalar ($\phi_H$) as the primary and lower branches, respectively.
In the present work, we do not display phantom scalarzied black holes. Their branch diagrams for horizon scalar and metric functions are given as Figs. 1 and 2 appeared in Ref.~\cite{Zou:2026staticthermo}.

\subsection{Radial perturbation scheme}
\label{radialnumerics}

We consider time-dependent radial perturbations around  the phantom scalarzied  black
holes~\cite{Blazquez-Salcedo:2018jnn,Blazquez-Salcedo:2022omw},
\begin{eqnarray}
&&\ud s^2=-A(r)e^{\eta F_t(t,r)}\ud t^2
+\frac{e^{\eta F_r(t,r)}}{B(r)}\ud r^2+r^2\ud\Omega_2^2,
\label{radialansatz}\\
&&\phi(t,r)=\phi_0(r)+\eta\varphi_1(t,r),
\label{radialexpansion}
\end{eqnarray}
where $\eta$ is an infinitesimal parameter, while $A$, $B$ and $\phi_0$ are the
background functions. $F_t$, $F_r$ and $\varphi_1$ describe the
perturbations.

Substituting into Eqs.~\eqref{Einstein} and \eqref{KG} and
retaining terms linear in $\eta$ gives the coupled perturbation equations.
Then, the linearized $tr$ and $rr$ equations allow us to eliminate the metric
perturbations, leaving a single equation for the scalar perturbation,
\begin{eqnarray}
&&g^2(r)\partial_t^2\varphi_1-\partial_r^2\varphi_1
+C_1(r)\partial_r\varphi_1+U(r)\varphi_1=0.
\label{radialtimemaster}
\end{eqnarray}
The functions of $g$, $C_1$ and $U$ depend only on the background and include
the metric response. Their derivation and explicit expressions are given in
Appendix~\ref{radialderivation}. In the spherically symmetric background, the scalar
perturbation is regarded as  the only dynamical mode.

\begin{samepage}
For a mode of frequency $\omega$, we write
\begin{eqnarray}
&&\varphi_1(t,r)=\chi(r)e^{-i\omega t}
\label{radialharmonic}
\end{eqnarray}
and obtain the radial equation
\begin{eqnarray}
&&\chi''-C_1\chi'+(\omega^2g^2-U)\chi=0.
\label{radialmaster}
\end{eqnarray}
Here a prime denotes $\ud/\ud r$.
\end{samepage}

For $g^2>0$ and regular metric constraints,
we introduce the effective tortoise coordinate $\rho$ and the radial wave function $\Psi_{\rm rad}$ through
\begin{eqnarray}
&&\frac{\ud\rho}{\ud r}=g,\qquad
\chi=C_0\Psi_{\rm rad},\qquad
\frac{C_0'}{C_0}=\frac12\left(C_1-\frac{g'}g\right).
\label{radialtransform}
\end{eqnarray}
We choose $g>0$ and the arbitrary normalization $C_0>0$. The last relation
removes the first derivative of $\Psi_{\rm rad}$. With this transformation,
Eq.~\eqref{radialmaster} takes the Schrödinger-type with potential $V_{\rm rad}$
\begin{eqnarray}
&&-\frac{\ud^2\Psi_{\rm rad}}{\ud\rho^2}+V_{\rm rad}\Psi_{\rm rad}
=\omega^2\Psi_{\rm rad},
\label{radialschrodinger}\\
&&V_{\rm rad}=\frac{1}{g^2}\left(U+C_1\frac{C_0'}{C_0}-\frac{C_0''}{C_0}\right)
\nonumber\\
&&\hspace{12mm}=\frac{U-C_1'/2+C_1^2/4}{g^2}
+\frac{g''}{2g^3}-\frac{3g'^2}{4g^4}.
\label{radialpotential}
\end{eqnarray}
The coupling generally makes $g$ being different from $(AB)^{-1/2}$ and the radial characteristic speed relative to the metric light cone is given by 
\begin{eqnarray}
&&c_{\rm rad}^2=\frac{1}{ABg^2}.
\label{radialspeed}
\end{eqnarray}
We note that the radial equation is hyperbolic for $c_{\rm rad}^2>0$  and its spectrum
determines the radial stability.

For the bald solution with $\zeta''(0)=0$, the above expressions reduce to those for Schwarzschild solution as
\begin{eqnarray}
&&g=\left(1-\frac{2M}{r}\right)^{-1},\qquad C_0\propto r^{-1},\qquad
V_{\rm rad}=\frac{2M}{r^3}\left(1-\frac{2M}{r}\right).
\label{radialbaldlimit}
\end{eqnarray}
Since this potential is positive,  we have   no
growing spherical scalar modes propagating around Schwarzschild black hole.

With the convention $e^{-i\omega t}$ in Eq.~\eqref{radialharmonic}, the ingoing condition at the future horizon and the outgoing condition at spatial infinity are
\begin{eqnarray}
&&\Psi_{\rm rad}\sim e^{-i\omega\rho}\quad(\rho\to-\infty),\qquad
\Psi_{\rm rad}\sim e^{i\omega\rho}\quad(\rho\to+\infty).
\label{radialQNMbc}
\end{eqnarray}
For an unstable perturbation with $\operatorname{Im}\omega>0$, these
conditions give a mode that decays at both ends of the tortoise line.
Multiplying Eq.~\eqref{radialschrodinger} by $\Psi_{\rm rad}^*$ and
integrating it gives
\begin{eqnarray}
&&\omega^2\int_{-\infty}^{\infty}|\Psi_{\rm rad}|^2\ud\rho
=\int_{-\infty}^{\infty}\left(|\partial_\rho\Psi_{\rm rad}|^2
+V_{\rm rad}|\Psi_{\rm rad}|^2\right)\ud\rho.
\label{radialenergyidentity}
\end{eqnarray}
Since the right-hand side is real, a mode with $\operatorname{Im}\omega>0$ must
have $\operatorname{Re}\omega=0$. Unstable modes thus correspond to
$\omega=i\Omega$ with $\Omega>0$, that is, to negative eigenvalues
$\omega^2=-\Omega^2$ of the real radial operator. For such modes, the boundary
conditions become
\begin{eqnarray}
&&\Psi_{\rm rad}\sim e^{\Omega\rho}\quad(\rho\to-\infty),\qquad
\Psi_{\rm rad}\sim e^{-\Omega\rho}\quad(\rho\to+\infty).
\label{radialboundstatebc}
\end{eqnarray}

For the regular nonextremal background, the near-horizon behavior is given by
\begin{eqnarray}
&&\rho\sim\frac{1}{2\kappa}\log\frac{r-r_H}{r_H},\qquad
\chi\propto(r-r_H)^{\Omega/(2\kappa)}\quad(r\to r_H),
\label{radialhorizonbc}
\end{eqnarray}
where $\kappa=\sqrt{a_1b_1}/2$ is the surface gravity  with $a_1$ and $b_1$
appeared in Eq.~\eqref{horizonexp}.

At spatial infinity, the tortoise coordinate and scalar amplitude behave as
\begin{eqnarray}
&&\rho=r+2M\log\frac{r}{r_H}+O(1),\qquad
\chi\propto\frac{e^{-\Omega\rho}}{r}\quad(r\to\infty).
\label{radialinfinitybc}
\end{eqnarray}
The corresponding limits of the effective potential are
\begin{eqnarray}
&&V_{\rm rad}\to0\quad(r\to r_H),\qquad
V_{\rm rad}=\frac{2M}{r^3}+O(r^{-4})\quad(r\to\infty).
\label{radialpotentialasymptotics}
\end{eqnarray}
An unstable mode is thus a normalizable bound state.

For a real, integrable potential on the full tortoise line, a useful sufficient conditions for instability is given by~\cite{Blazquez-Salcedo:2018jnn,Myung:2019radialcomparison}
\begin{eqnarray}
&&I\equiv\int_{-\infty}^{\infty}V_{\rm rad}\ud\rho
=\int_{r_H}^{\infty}V_{\rm rad}g\ud r,\qquad \widehat I=r_H I,
\label{radialintegralcriterion}\\
&&V_{\rm rad}\geq0\ \Longrightarrow\ \omega^2\geq0,\qquad
I<0\ \Longrightarrow\ \text{at least one unstable mode}.
\label{radialsufficientconditions}
\end{eqnarray}
We wish to compute the radial spectrum directly to determine the number of negative
eigenvalues and their growth rates.
To resolve the region near the horizon, we integrate the radial equation by introducing 
the coordinate $z=\log(r/r_H-1)$. In this coordinate,
Eq.~\eqref{radialmaster} reads as 
\begin{eqnarray}
\chi_{zz}+a\chi_z+[(\omega r_H)^2b+c]\chi=0,
\label{radiallog}
\end{eqnarray}
where
\begin{eqnarray}
a=-1-r_He^zC_1,\qquad b=e^{2z}g^2,\qquad c=-r_H^2e^{2z}U.
\label{radiallogcoefficients}
\end{eqnarray}

Multiplying Eq.(\ref{radiallog})  by $\mu=\exp(\int a\ud z)$ gives the self-adjoint form
\begin{eqnarray}
-(\mu\chi_z)_z-\mu c\chi=(\omega r_H)^2\mu b\chi.
\label{radialselfadjoint}
\end{eqnarray}
We then discretize Eq.(\ref{radialselfadjoint})  as a symmetric finite-volume eigenvalue problem, with
$\chi=0$ at $z=-50$ and at $r=300r_H$, or at $r=400r_H$ for the canonical
polynomial solutions.  The discretization reproduces the
Pöschl--Teller spectrum and the Schwarzschild limit. The number of
negative eigenvalues does not change  when the grid is refined  and either  when the
horizon cutoff is varied between $z=-40$ and $-80$ or when the outer boundary
is placed between $150r_H$ and $800r_H$.  For the exponential lower branch,
the grid is doubled from 4000 to 8000 points.

For growing modes, we determine the growth rate independently with the
logarithmic derivative $u=\chi_z/\chi$,
\begin{eqnarray}
&&u_z=-u^2-au+(\Omega r_H)^2b-c,\qquad
u_L(z_m;\Omega)-u_R(z_m;\Omega)=0.
\label{radialmatching}
\end{eqnarray}
Here, the left and right solutions satisfy Eqs.~\eqref{radialhorizonbc} and
\eqref{radialinfinitybc}, respectively, and their matching at $z_m$ fixes
$\Omega$. The spectral and matching growth rates differ by less than
$2\times10^{-6}$ for the polynomial lower branch and $4\times10^{-6}$
for the exponential lower branch.  From the growing eigenfunctions, we also
evaluate the Rayleigh quotient defined by 
\begin{eqnarray}
&&{\cal R}=\frac{\displaystyle\int\left(|\partial_\rho\Psi_{\rm rad}|^2
+V_{\rm rad}|\Psi_{\rm rad}|^2\right)\ud\rho}
{\displaystyle\int|\Psi_{\rm rad}|^2\ud\rho}
\label{radialRayleighnumerics}
\end{eqnarray}
for comparison with $-\Omega^2$.

\section{Radial stability of phantom scalarized black holes}
\label{radialstability}

\subsection{Effective potentials and unstable radial modes}
\label{lowerradialstability}

\begin{figure}[htbp]
\centering
\subfigure[$\zeta_p,\ \beta_p=125$]{\label{figradialpotential:zeta1}
\includegraphics[width=0.315\textwidth]{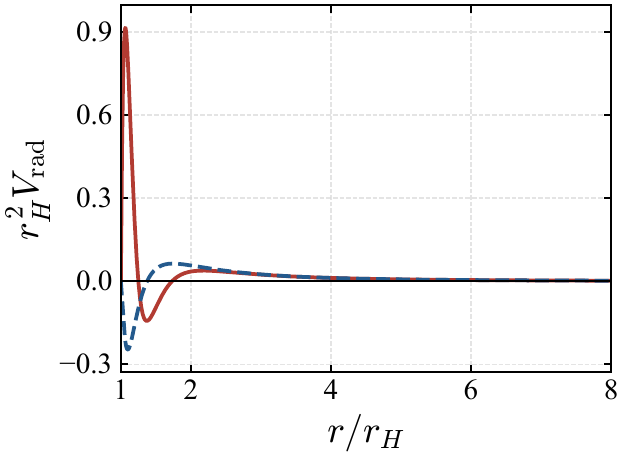}}
\hfill
\subfigure[$\zeta_e,\ \beta_e=1000$]{\label{figradialpotential:zetae}
\includegraphics[width=0.315\textwidth]{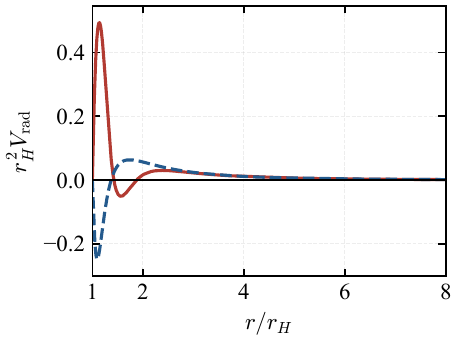}}
\hfill
\subfigure[$\zeta_p,\ \beta_p=0$]{\label{figradialpotential:zeta3}
\includegraphics[width=0.315\textwidth]{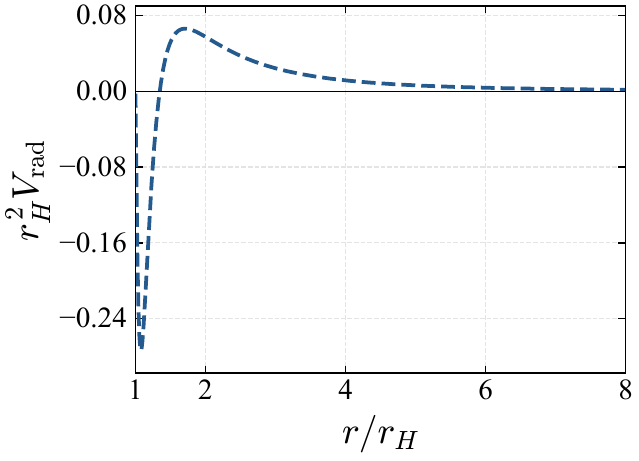}}
\caption{Effective radial potentials at $r_H=0.10$. Red solid: primary and  blue dashed: lower. The coupling and parameter are indicated below each panel.}\label{figradialpotential}
\end{figure}

Figure~\ref{figradialpotential} shows the effective potentials at $r_H=0.10$
for the polynomial coupling with $\alpha=1/4$ and $\beta_p=125$ or $0$, and
for the exponential coupling with $\beta_e=1000$. The red solid and blue
dashed curves denote the primary and lower branches, respectively and  the pure
quartic solutions lie on the lower branch. Each lower-branch potential has a
negative well near the horizon and $\widehat I<0$. Solving the radial
equation gives  one positive $\Omega$ with negative $\widehat{I}$ for each lower solution.
Table~\ref{tabradial} lists representative backgrounds and their radial
results. At the same $r_H$, the primary and lower backgrounds have
different ADM masses.

\begin{table}[htbp]
\centering\small
\setlength{\tabcolsep}{4pt}
\begin{tabular}{lllrrrrr}
\toprule
Coupling & Parameter & Branch & $r_H/\lambda$ & $\phi_H$ & $M/\lambda$ & $M\Omega$ & $\widehat I$\\
\midrule
$\zeta_p$ & $\beta_p=125$ & primary & 0.100 & 0.175781 & 0.0487562 & --- & 2.4150\\
$\zeta_p$ & $\beta_p=125$ & lower & 0.100 & 0.1216 & 0.049419 & 0.1054 & $-0.3857$\\
\midrule
$\zeta_e$ & $\beta_e=1000$ & primary & 0.100 & 0.232446 & 0.0478622 & --- & 1.3167\\
$\zeta_e$ & $\beta_e=1000$ & lower & 0.100 & 0.121200 & 0.0494210 & 0.10819 & $-0.4038$\\
\midrule
$\zeta_p$ & $\beta_p=0$ & lower & 0.100 & 0.116305 & 0.0494537 & 0.1242 & $-0.4984$\\
\bottomrule
\end{tabular}
\caption{Representative radial modes and potential integrals. A large dash denotes that no growing mode is found.}\label{tabradial}
\end{table}

\FloatBarrier

\begin{figure}[htbp]
\centering
\subfigure[$\zeta_p,\ \beta_p=125$]{\label{figradialgrowth:zeta1}
\includegraphics[width=0.315\textwidth]{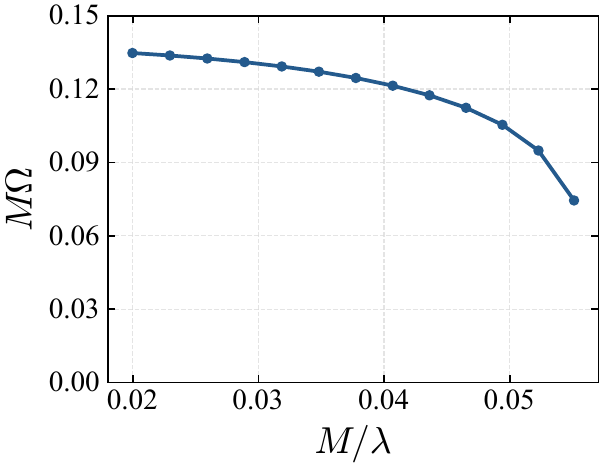}}
\hfill
\subfigure[$\zeta_e,\ \beta_e=1000$]{\label{figradialgrowth:zetae}
\includegraphics[width=0.315\textwidth]{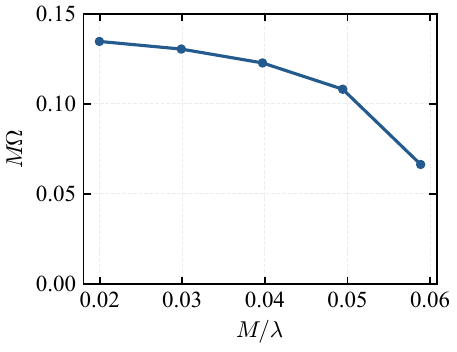}}
\hfill
\subfigure[$\zeta_p,\ \beta_p=0$]{\label{figradialgrowth:zeta3}
\includegraphics[width=0.315\textwidth]{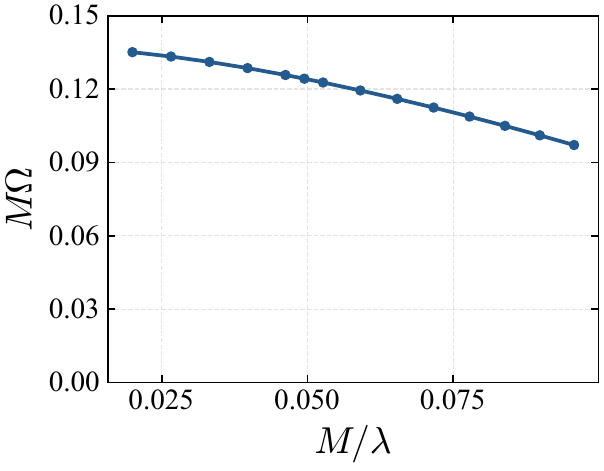}}
\caption{Lower-branch growth rates for different  coupling and parameter indicated below each panel.}\label{figradialgrowth}
\end{figure}

Figure~\ref{figradialgrowth} shows lower-branch growth rates in the range
$0.06638\lesssim M\Omega\lesssim0.1351$, with a maximum of $0.13473$ for
the exponential coupling. The corresponding e-folding times
$\tau=1/\Omega$ therefore lie in the range
$7.40\lesssim\tau/M\lesssim15.07$. At $\beta_p=125$ and
$\beta_e=1000$, the growth slows towards larger masses.

The growing eigenfunctions are localized and nodeless, as expected for the
lowest bound state.
For the lower solutions of Table~\ref{tabradial}, at $\beta_p=125$,
$\beta_e=1000$ and $\beta_p=0$, the Rayleigh quotient~\eqref{radialRayleighnumerics}
gives $r_H^2\mathcal R\simeq-0.04546,-0.04792,-0.06311$, respectively, in good 
agreement with $-(\Omega r_H)^2$.
Moreover, the metric constraints remain regular outside the horizon. The radial
characteristic speed is positive, with a minimum
$c_{\rm rad}^2\simeq0.9427$ for the polynomial coupling and
$c_{\rm rad}^2\geq0.9781$ for the exponential coupling, so these growing modes
appear  in the hyperbolic regime.

\FloatBarrier

\subsection{Zero-energy solutions and \texorpdfstring{$S$}{S}-deformation method}
\label{primaryradialstability}

\begin{samepage}
For both couplings, the primary-branch spectrum has no negative eigenvalue
and $0.4870\lesssim\widehat I\lesssim6.6932$. Since the potentials in
Figs.~\ref{figradialpotential:zeta1} and \ref{figradialpotential:zetae}
show changing sign, we use the $S$-deformation method~\cite{Kimura:2018whv},
starting from the horizon-regular zero-energy solution  as 
\begin{eqnarray}
&&{\cal H}\psi_0=0,\qquad
{\cal H}=-\frac{\ud^2}{\ud\rho^2}+V_{\rm rad},\qquad
\psi_0\to1,\quad\partial_\rho\psi_0\to0\quad(\rho\to-\infty).
\label{radialzerocriterion}
\end{eqnarray}
\end{samepage}
If $\psi_0$ remains strictly positive throughout the exterior,
${\cal S}=-\partial_\rho\log\psi_0$ is real and has no pole. One finds 
\begin{eqnarray}
&&V_{\rm rad}+\partial_\rho{\cal S}-{\cal S}^2=0,\qquad
{\cal H}=(-\partial_\rho+{\cal S})(\partial_\rho+{\cal S}).
\label{radialfactorization}
\end{eqnarray}
For compactly supported test functions, or normalizable growing modes with
vanishing boundary terms, the factorization leads to 
\begin{eqnarray}
&&\int_{-\infty}^{\infty}\left(|\partial_\rho\Psi|^2+V_{\rm rad}|\Psi|^2\right)\ud\rho
=\int_{-\infty}^{\infty}|\partial_\rho\Psi+{\cal S}\Psi|^2\ud\rho\geq0.
\label{radialpositiveform}
\end{eqnarray}
Together with Eq.~\eqref{radialenergyidentity}, this gives a nonnegative
radial spectrum.

We then integrate the auxiliary equation for the scalar amplitude
\begin{eqnarray}
&&\chi_0''-C_1\chi_0'-U\chi_0=0.
\label{radialzeroequation}
\end{eqnarray}
The horizon-regular solution is normalized by
\begin{eqnarray}
&&\chi_0\to1,\quad\partial_\rho\chi_0\to0\quad(r\to r_H).
\label{radialzerohorizonbc}
\end{eqnarray}
At spatial infinity, the asymptotic solution is expressed as 
\begin{eqnarray}
&&\chi_0=A_\infty+B_\infty\frac{r_H}{r}
+O\!\left(\frac{r_H^2}{r^2}\right).
\label{radialzeroasymptotics}
\end{eqnarray}
Since $\chi_0=C_0\psi_0$ with $C_0>0$, positivity of $\chi_0$ implies
positivity of $\psi_0$. Table~\ref{tabprimarystability} shows that $\chi_0$
remains positive and it approaches a positive constant $A_\infty$, fitted with
Eq.~\eqref{radialzeroasymptotics}. The exponential solutions are integrated
to $400r_H$ and  a Prüfer integration confirms the absence of nodes for the
polynomial solutions. The $S$-deformation therefore establishes radial
stability for these primary branch. In contrast, the lower-branch
zero-energy solutions have one node. The background scalar profiles are
nodeless on both branches.

\begin{table}[htbp]
\centering\small
\caption{Primary-branch zero-energy solutions at $\beta_p=125$ for $\zeta_p$ and $\beta_e=1000$ for $\zeta_e$.}\label{tabprimarystability}
\setlength{\tabcolsep}{7pt}
\begin{tabular}{lrrrr}
\toprule
& $r_H/\lambda$ & $\widehat I$ & $\min\chi_0$ & $A_\infty$\\
\midrule
$\zeta_p$ & 0.08 & 6.6932 & 1.00000 & 213.15658\\
$\zeta_p$ & 0.10 & 2.4150 & 1.00000 & 13.70178\\
$\zeta_p$ & 0.112 & 0.7932 & 1.00000 & 1.87893\\
\midrule
$\zeta_e$ & 0.04 & 3.7660 & 1.00000 & 44.79824\\
$\zeta_e$ & 0.06 & 2.8879 & 1.00000 & 20.22266\\
$\zeta_e$ & 0.08 & 2.0809 & 1.00000 & 9.70941\\
$\zeta_e$ & 0.10 & 1.3167 & 1.00000 & 4.26150\\
$\zeta_e$ & 0.12 & 0.4870 & 0.99825 & 0.99640\\
\bottomrule
\end{tabular}
\end{table}

\FloatBarrier

The stable exponential primary branch  implies  that a finite nonzero root of
$\zeta_e'$ is unnecessary for radial stability, in agreement with the
canonical exponential results of Ref.~\cite{Blazquez-Salcedo:2022omw}.
For $\beta_e=1000$, that study also found a loss of radial hyperbolicity at
small masses. By contrast, the phantom primary branch at the same
$\beta_e$ retains a positive characteristic speed down to $r_H=0.04$.

\FloatBarrier

\subsection{Canonical polynomial black holes}
\label{canonicalcomparison}

We next compute the radial modes of the canonical polynomial black holes
constructed in Ref.~\cite{Zou:2024paper}. At $\lambda=1$ and $\alpha=1/4$,
we compare the $\beta_p=0$ and $125$ solutions with the phantom ones at the
same horizon radii $r_H=0.08$ and $0.10$. Their ADM masses differ between
the two signs. For the canonical solutions, we use the same reduction
and boundary conditions as for the phantom ones. The constraints again remain
regular, and the characteristic speed is positive outside the horizon.

As is shown in Table~\ref{tabsigncomparison}, each lower solution has one growing
mode for both signs ($\epsilon=\mp1$). Here, $N_\epsilon$ counts growing radial modes, and a
large dash denotes their absence. The canonical primary solutions are radially stable, with
positive zero-energy solutions:  $\min\chi_0=1$, and $A_\infty=245.04010$ and
$16.67830$ at $r_H=0.08$ and $0.10$, respectively. At the same couplings and
horizon radii, the canonical values of $M\Omega$ exceed the phantom ones.
The backgrounds and metric constraints depend on $\epsilon$ and therefore, they
lead to different coupled radial spectra.

\begin{table}[htbp]
\centering\small
\caption{Matched polynomial backgrounds for $\epsilon=-1$ (phantom) and $+1$ (canonical).}
\label{tabsigncomparison}
\setlength{\tabcolsep}{6pt}
\begin{tabular}{llrrrrr}
\toprule
$\beta_p$ & Branch & $r_H/\lambda$ & $N_{-1}$ & $N_{+1}$ & $(M\Omega)_{-1}$ & $(M\Omega)_{+1}$\\
\midrule
$0$ & lower & 0.08 & 1 & 1 & 0.12856 & 0.14961\\
$0$ & lower & 0.10 & 1 & 1 & 0.12423 & 0.16033\\
\midrule
$125$ & lower & 0.08 & 1 & 1 & 0.12249 & 0.14217\\
$125$ & lower & 0.10 & 1 & 1 & 0.10537 & 0.13323\\
$125$ & primary & 0.08 & 0 & 0 & --- & ---\\
$125$ & primary & 0.10 & 0 & 0 & --- & ---\\
\bottomrule
\end{tabular}
\end{table}

\FloatBarrier

\subsection{Comparison with thermodynamic analysis}
\label{thermocomparison}
It is important to develop the connection between dynamical stability and thermodynamic analysis.
The Wald entropy of the scalarized black holes is given
by~\cite{Zou:2026staticthermo}
\begin{eqnarray}
&&S_H=\pi r_H^2+4\pi\epsilon\lambda^2\zeta(\phi_H).
\label{thermo}
\end{eqnarray}
To compare with Schwarzschild black holes, we define the entropy and
free-energy differences at fixed mass and temperature, respectively,
\begin{eqnarray}
&&\delta S=S_H-4\pi M^2,\qquad \delta F=F-\frac{1}{16\pi T_H}.
\label{thermodifferences}
\end{eqnarray}
Here, $T_H=\kappa/2\pi$ is the Hawking temperature and $F=M-T_HS_H$ is the
free energy. A solution with $\delta S>0$ has larger entropy than the
Schwarzschild black hole with the same mass and a solution with $\delta F<0$
is thermodynamically preferred over the Schwarzschild black hole at the same
temperature.  As was suggested in Ref.~\cite{Zou:2026staticthermo}, we label the solutions
by
\begin{equation}
x=\begin{cases}
(8\beta_p)^{1/4}\phi_H, & \text{polynomial coupling with }\beta_p=125,\\
\beta_e^{1/4}\phi_H, & \text{exponential coupling},
\end{cases}
\label{eq:poly-parameter}
\end{equation}
For the polynomial coupling, the nonzero root of $\zeta_p'$ corresponds to
$x=1$.

For both polynomial and exponential couplings,
Ref.~\cite{Zou:2026staticthermo} has shown  that the phantom lower branches and
portions of the primary branches are thermodynamically preferred over
Schwarzschild black holes at the same temperature.
The phantom lower solutions in Table~\ref{tabradial} and
Fig.~\ref{figradialgrowth} have $\delta S>0$ and $\delta F<0$ with one growing
radial mode.
The canonical polynomial lower solutions appeared in
Table~\ref{tabsigncomparison} have $\delta S<0$ and $\delta F>0$ with one growing mode
as well.

On the phantom primary branches, the stable solutions lie
on both sides of the free-energy crossing $x_F$ found in
Ref.~\cite{Zou:2026staticthermo}. For $\beta_p=125$, the solutions with
$r_H=0.08$ and $0.10$ have $x>x_F$ and $\delta F>0$, and the solution with
$r_H=0.112$ has $\delta F<0$. For $\beta_e=1000$, we find $\delta F>0$ for
$0.04\leq r_H\leq0.10$ and $\delta F<0$ at $r_H=0.12$. We find that the canonical polynomial
primary solutions at $r_H=0.08$ and $0.10$ have $\delta F<0$.

For the canonical polynomial background, it is clear that  the free-energy ordering agrees with the
radial stability.  But  it does not work for the phantom  polynomial background.
For the canonical
exponential model, Ref.~\cite{Blazquez-Salcedo:2022omw} likewise found that
the radially stable branch is thermodynamically preferred, in agreement with
Ref.~\cite{Doneva:2021tvn}.  The backgrounds used in our radial calculations
lie outside the narrow regions of positive heat capacity between the
temperature minimum $x_T$ and the mass maximum $x_M$
\cite{Zou:2026staticthermo}. However, since  Schwarzschild black hole has 
entirely negative heat capacity,  its heat capacity does not distinguish the stable
and unstable  black hole solutions either.

Finally, the radial results also indicate the change of stability along the phantom
families. For $\beta_p=125$, the number of growing modes changes from one to
zero between the lower and primary solutions with $r_H=0.112$, which have
$x\simeq0.804$ and $0.956$. This range contains the mass maximum
$x_M\simeq0.89079$ of the phantom family \cite{Zou:2026staticthermo}. For
$\beta_e=1000$, the solutions with $r_H=0.12$ have $x\simeq0.886$ and
$1.152$, and the range contains $x_M\simeq1.0142$. Towards these solutions,
$M\Omega$ decreases on the lower branch (see Fig.~\ref{figradialgrowth}) and
$A_\infty$ decreases on the primary branch (see Table~\ref{tabprimarystability}).
For the canonical exponential coupling, Ref.~\cite{Blazquez-Salcedo:2022omw}
found a zero mode at the maximal mass, separating the stable branch from the unstable
one. A calculation near the phantom turning points would determine
the corresponding static modes.

\FloatBarrier

\section{Conclusions and discussions}
\label{con}

We have analyzed the radial stability of the phantom scalarized black holes
constructed in Ref.~\cite{Zou:2026staticthermo} by reducing the coupled
scalar--metric perturbations to a Schr\"odinger-type equation with an effective
potential. Two types of coupling are employed to realize the nonlinear
scalarization, namely the polynomial and the exponential ones.

We first consider the polynomial coupling. Although both the lower and the
primary branches of the phantom family possess nodeless scalar profiles, their
stability properties differ: the lower branch has one growing mode, whereas the
primary branch admits positive zero-energy perturbations and is therefore stable
according to the $S$-deformation method. The pure quartic case is unstable as
well.
For the canonical theory, our calculation reproduces this pattern for both
couplings, with an unstable lower and a stable primary branch. Moreover, the
exponential branches show that a finite nonzero root of the coupling derivative
is not sufficient for radial stability, in agreement with
Ref.~\cite{Blazquez-Salcedo:2022omw}. At fixed couplings and horizon radii,
$M\Omega$ of the canonical lower branch is larger than that of the phantom one,
indicating that the difference in metric backreaction modifies the timescale of
the instability.

As for the thermodynamic stability, Ref.~\cite{Zou:2026staticthermo} found from
the Helmholtz free energy that, at fixed temperature, the phantom lower branch
and part of the primary branch are thermodynamically preferred over the bald
Schwarzschild black hole. This free-energy ordering agrees with the radial
stability of the canonical polynomial and exponential branches, but not with
that of the phantom branches, so that the two criteria are only partially
consistent.

The stability boundary can be located by computing the zero mode near the
turning point of the phantom families, where the mass reaches its maximum.
Extending the exponential families to smaller masses, and to $\beta_e=100$ and
$25$, would test whether they lose hyperbolicity of the radial perturbation
equations as in the canonical theory~\cite{Blazquez-Salcedo:2022omw}. A study of
non-spherical perturbations and nonlinear evolution would further clarify the
stability and the dynamical fate of the phantom branches.

\FloatBarrier

\section*{Acknowledgments}

This research is supported by the Jiangxi Provincial Natural Science
Foundation (No.~20262BAC240347), the National Natural Science Foundation of
China (NNSFC) (No.~12365009), and the Natural Science Basic Research Program
of Shaanxi Province (Program No.~2026JC-YBMS-0083).

\appendix
\numberwithin{equation}{section}

\section{Radial perturbation equations}
\label{radialderivation}

For the radial perturbation analysis in Sec.~\ref{radialnumerics}, we write
$F_t=-H_0(r)e^{-i\omega t}$ and $F_r=H_2(r)e^{-i\omega t}$ together  with the
scalar amplitude $\chi(r)$ defined in Eq.~\eqref{radialharmonic}.

For $\omega\ne0$,  the metric
perturbations can be expressed  in terms of the scalar amplitude and its first derivative as 
\begin{eqnarray}
&&H_2=h_0\chi+h_1\chi',\qquad
H_0'=(k_0+\omega^2k_\omega)\chi+k_1\chi'.
\label{radialreconstruction}
\end{eqnarray}
The coefficients depend only on the phantom background. To write them
explicitly, we suppress the subscript in $\phi_0$ and define new quantities 
\begin{eqnarray}
&&Z_j=\lambda^2\frac{\ud^j\zeta}{\ud\phi^j},\qquad
D=r+2\epsilon(1-3B)Z_1\phi'.
\label{radialconstraint}
\end{eqnarray}
Primes on $A$, $B$ and $\phi$ denote $\ud/\ud r$, so that one has 
$Z_j'=Z_{j+1}\phi'$. The metric coefficients in
Eq.~\eqref{radialreconstruction} are given by 
\begin{eqnarray}
&&h_0=-\frac{2\epsilon[(1-B)(Z_1A'/A-2Z_2\phi')-r^2\phi']}D,\qquad
h_1=-\frac{4\epsilon(B-1)Z_1}D,
\label{radialmetriccoeffa}\\
&&k_0=-\frac{1-6\epsilon B^2Z_1\phi'A'/A}{BD}h_0
-\frac{2\epsilon(3B-1)Z_2\phi'A'}{AD},
\nonumber\\
&&k_1=-\frac{1-6\epsilon B^2Z_1\phi'A'/A}{BD}h_1
-\frac{2\epsilon[(3B-1)Z_1A'/A+r^2\phi']}D,
\nonumber\\
&&k_\omega=-\frac{4\epsilon(B-1)Z_1}{ABD}.
\label{radialmetriccoeffb}
\end{eqnarray}
The linearized scalar equation takes a complicated  form
\begin{eqnarray}
&&0=B\chi''+\left(\frac{2B}{r}+\frac{BA'}{2A}+\frac{B'}2\right)\chi'
+\left(\frac{\omega^2}{A}+\frac{Z_2{\cal G}_0}{4}\right)\chi
\nonumber\\
&&\hspace{7mm}+s_2H_0''+s_tH_0'+s_rH_2'+(s_0+\omega^2s_\omega)H_2.
\label{radiallinearscalar}
\end{eqnarray}
Here, ${\cal G}_0$ is the Gauss--Bonnet invariant of the phantom background as
\begin{eqnarray}
&&{\cal G}_0=\frac{2}{r^2}\left[(3B-1)B'\frac{A'}A
+B(B-1)\left(\frac{2A''}A-\frac{A'^2}{A^2}\right)\right].
\label{radialbackgroundcoefficients}
\end{eqnarray}
The coefficients for the metric perturbations are found to be
\begin{eqnarray}
&&s_2=\frac{B(1-B)Z_1}{r^2},
\nonumber\\
&&s_t=\frac{Z_1[(1-3B)B'+2B(1-B)A'/A]-Br^2\phi'}{2r^2},
\nonumber\\
&&s_r=\frac{B[(1-3B)Z_1A'/A-r^2\phi']}{2r^2},
\nonumber\\
&&s_0=-B\phi''-\left(\frac{2B}{r}+\frac{BA'}{2A}+\frac{B'}2\right)\phi'
\nonumber\\
&&\hspace{9mm}-\frac{Z_1}{2r^2}\left[(6B-1)B'\frac{A'}A
+B(2B-1)\left(\frac{2A''}A-\frac{A'^2}{A^2}\right)\right],
\nonumber\\
&&s_\omega=\frac{(B-1)Z_1}{Ar^2}.
\label{radialscalarcoefficients}
\end{eqnarray}

Eliminating $H_0$ and $H_2$ with Eq.~\eqref{radialreconstruction} leads to  a
single equation for $\chi$,
\begin{eqnarray}
&&p_2\chi''+p_1\chi'+(\omega^2p_\omega+p_0)\chi=0,\qquad
g^2=\frac{p_\omega}{p_2},\quad C_1=-\frac{p_1}{p_2},\quad U=-\frac{p_0}{p_2},
\label{radialcoefficients}
\end{eqnarray}
with four $p_i$ as 
\begin{eqnarray}
&&p_2=B+s_2k_1+s_rh_1,
\nonumber\\
&&p_1=\frac{2B}{r}+\frac{BA'}{2A}+\frac{B'}2
+s_2(k_0+k_1')+s_tk_1+s_r(h_0+h_1')+s_0h_1,
\nonumber\\
&&p_\omega=\frac1A+s_2k_\omega'+s_tk_\omega+s_\omega h_0,
\nonumber\\
&&p_0=\frac{Z_2{\cal G}_0}{4}+s_2k_0'+s_tk_0+s_rh_0'+s_0h_0.
\label{radialexplicitcoefficients}
\end{eqnarray}
The terms proportional to $\omega^2\chi'$ disappears  because of the relation 
$s_2k_\omega+s_\omega h_1=0$. For nonsingular metric constraints and
$p_2\ne0$, the definitions of $g^2$, $C_1$ and $U$ in
Eq.~\eqref{radialcoefficients} supply  Eq.~\eqref{radialmaster} and the
time-dependent equation~\eqref{radialtimemaster}.   At $\omega=0$, a physical
static mode reduces to  the original metric constraint.

For the exponential coupling, the derivatives of $\zeta_e$ with respect to
$\phi$ are given by 
\begin{eqnarray}
&&\zeta_e'=\phi^3e^{-\beta_e\phi^4},\qquad
\zeta_e''=\phi^2(3-4\beta_e\phi^4)e^{-\beta_e\phi^4},
\nonumber\\
&&\zeta_e'''=\phi(6-36\beta_e\phi^4+16\beta_e^2\phi^8)e^{-\beta_e\phi^4}.
\label{exponentialderivatives}
\end{eqnarray}

\FloatBarrier

\begingroup

\linespread{1.3}
\small\selectfont
\let\originalthebibliography\thebibliography
\renewcommand{\thebibliography}[1]{%
  \originalthebibliography{#1}%
  \setlength{\itemsep}{3pt}%
  \setlength{\parsep}{0pt}%
  \setlength{\parskip}{0pt}%
}

\bibliographystyle{unsrt}

\bibliography{references}

@article{Damour:1993hw,
  author = {T.~Damour and G.~Esposito-Far\`ese},
  title = {{Nonperturbative strong-field effects in tensor-scalar theories of gravitation}},
  journal = {Phys. Rev. Lett.},
  volume = {70},
  pages = {2220},
  year = {1993},
}

@article{Bekenstein:1974sf,
  author = {J.~D.~Bekenstein},
  title = {{Exact solutions of Einstein conformal scalar equations}},
  journal = {Annals Phys.},
  volume = {82},
  pages = {535},
  year = {1974},
}

@article{Bekenstein:1975ts,
  author = {J.~D.~Bekenstein},
  title = {{Black holes with scalar charge}},
  journal = {Annals Phys.},
  volume = {91},
  pages = {75},
  year = {1975},
}

@article{Bekenstein:1995un,
  author = {J.~D.~Bekenstein},
  title = {{Novel no-scalar-hair theorem for black holes}},
  journal = {Phys. Rev. D},
  volume = {51},
  pages = {R6608},
  year = {1995},
}

@article{Herdeiro:2015waa,
  author = {C.~A.~R.~Herdeiro and E.~Radu},
  title = {{Asymptotically flat black holes with scalar hair: a review}},
  journal = {Int. J. Mod. Phys. D},
  volume = {24},
  pages = {1542014},
  year = {2015},
  eprint = {1504.08209},
  archiveprefix = {arXiv},
  primaryclass = {gr-qc},
  note = {[arXiv:1504.08209 [gr-qc]]},
}

@article{Kanti:1995vq,
  author = {P.~Kanti and N.~E.~Mavromatos and J.~Rizos and K.~Tamvakis and E.~Winstanley},
  title = {{Dilatonic black holes in higher curvature string gravity}},
  journal = {Phys. Rev. D},
  volume = {54},
  pages = {5049--5058},
  year = {1996},
  eprint = {hep-th/9511071},
  archiveprefix = {arXiv},
  note = {[arXiv:hep-th/9511071]},
}

@article{Sotiriou:2013qea,
  author = {T.~P.~Sotiriou and S.~Y.~Zhou},
  title = {{Black hole hair in generalized scalar-tensor gravity}},
  journal = {Phys. Rev. Lett.},
  volume = {112},
  pages = {251102},
  year = {2014},
  eprint = {1312.3622},
  archiveprefix = {arXiv},
  primaryclass = {gr-qc},
  note = {[arXiv:1312.3622 [gr-qc]]},
}

@article{Doneva:2017bvd,
  author = {D.~D.~Doneva and S.~S.~Yazadjiev},
  title = {{New Gauss-Bonnet black holes with curvature-induced scalarization in extended scalar-tensor theories}},
  journal = {Phys. Rev. Lett.},
  volume = {120},
  pages = {131103},
  year = {2018},
  eprint = {1711.01187},
  archiveprefix = {arXiv},
  primaryclass = {gr-qc},
  note = {[arXiv:1711.01187 [gr-qc]]},
}

@article{Silva:2017uqg,
  author = {H.~O.~Silva and J.~Sakstein and L.~Gualtieri and T.~P.~Sotiriou and E.~Berti},
  title = {{Spontaneous scalarization of black holes and compact stars from a Gauss-Bonnet coupling}},
  journal = {Phys. Rev. Lett.},
  volume = {120},
  pages = {131104},
  year = {2018},
  eprint = {1711.02080},
  archiveprefix = {arXiv},
  primaryclass = {gr-qc},
  note = {[arXiv:1711.02080 [gr-qc]]},
}

@article{Antoniou:2017acq,
  author = {G.~Antoniou and A.~Bakopoulos and P.~Kanti},
  title = {{Evasion of no-hair theorems and novel black-hole solutions in Gauss-Bonnet theories}},
  journal = {Phys. Rev. Lett.},
  volume = {120},
  pages = {131102},
  year = {2018},
  eprint = {1711.03390},
  archiveprefix = {arXiv},
  primaryclass = {hep-th},
  note = {[arXiv:1711.03390 [hep-th]]},
}

@article{Antoniou:2017hxj,
  author = {G.~Antoniou and A.~Bakopoulos and P.~Kanti},
  title = {{Black-hole solutions with scalar hair in Einstein-scalar-Gauss-Bonnet theories}},
  journal = {Phys. Rev. D},
  volume = {97},
  pages = {084037},
  year = {2018},
  eprint = {1711.07431},
  archiveprefix = {arXiv},
  primaryclass = {hep-th},
  note = {[arXiv:1711.07431 [hep-th]]},
}

@article{Doneva:2022ewd,
  author = {D.~D.~Doneva and F.~M.~Ramazano\u{g}lu and H.~O.~Silva and T.~P.~Sotiriou and S.~S.~Yazadjiev},
  title = {{Spontaneous scalarization}},
  journal = {Rev. Mod. Phys.},
  volume = {96},
  pages = {015004},
  year = {2024},
  eprint = {2211.01766},
  archiveprefix = {arXiv},
  primaryclass = {gr-qc},
  note = {[arXiv:2211.01766 [gr-qc]]},
}

@article{Myung:2018iyq,
  author = {Y.~S.~Myung and D.~C.~Zou},
  title = {{Gregory-Laflamme instability of black hole in Einstein-scalar-Gauss-Bonnet theories}},
  journal = {Phys. Rev. D},
  volume = {98},
  pages = {024030},
  year = {2018},
  eprint = {1805.05023},
  archiveprefix = {arXiv},
  primaryclass = {gr-qc},
  note = {[arXiv:1805.05023 [gr-qc]]},
}

@article{Blazquez-Salcedo:2018jnn,
  author = {J.~L.~Bl\'azquez-Salcedo and D.~D.~Doneva and J.~Kunz and S.~S.~Yazadjiev},
  title = {{Radial perturbations of the scalarized Einstein-Gauss-Bonnet black holes}},
  journal = {Phys. Rev. D},
  volume = {98},
  pages = {084011},
  year = {2018},
  eprint = {1805.05755},
  archiveprefix = {arXiv},
  primaryclass = {gr-qc},
  note = {[arXiv:1805.05755 [gr-qc]]},
}

@article{Silva:2018qhn,
  author = {H.~O.~Silva and C.~F.~B.~Macedo and T.~P.~Sotiriou and L.~Gualtieri and J.~Sakstein and E.~Berti},
  title = {{Stability of scalarized black hole solutions in scalar-Gauss-Bonnet gravity}},
  journal = {Phys. Rev. D},
  volume = {99},
  pages = {064011},
  year = {2019},
  eprint = {1812.05590},
  archiveprefix = {arXiv},
  primaryclass = {gr-qc},
  note = {[arXiv:1812.05590 [gr-qc]]},
}

@article{Blazquez-Salcedo:2020rhf,
  author = {J.~L.~Bl\'azquez-Salcedo and D.~D.~Doneva and S.~Kahlen and J.~Kunz and P.~Nedkova and S.~S.~Yazadjiev},
  title = {{Axial perturbations of the scalarized Einstein-Gauss-Bonnet black holes}},
  journal = {Phys. Rev. D},
  volume = {101},
  pages = {104006},
  year = {2020},
  eprint = {2003.02862},
  archiveprefix = {arXiv},
  primaryclass = {gr-qc},
  note = {[arXiv:2003.02862 [gr-qc]]},
}

@article{Blazquez-Salcedo:2020caw,
  author = {J.~L.~Bl\'azquez-Salcedo and D.~D.~Doneva and S.~Kahlen and J.~Kunz and P.~Nedkova and S.~S.~Yazadjiev},
  title = {{Polar quasinormal modes of the scalarized Einstein-Gauss-Bonnet black holes}},
  journal = {Phys. Rev. D},
  volume = {102},
  pages = {024086},
  year = {2020},
  eprint = {2006.06006},
  archiveprefix = {arXiv},
  primaryclass = {gr-qc},
  note = {[arXiv:2006.06006 [gr-qc]]},
}

@article{Minamitsuji:2024twp,
  author = {M.~Minamitsuji and S.~Mukohyama and S.~Tsujikawa},
  title = {{Angular and radial stabilities of spontaneously scalarized black holes in the presence of scalar-Gauss-Bonnet couplings}},
  journal = {Phys. Rev. D},
  volume = {109},
  pages = {104057},
  year = {2024},
  eprint = {2403.10048},
  archiveprefix = {arXiv},
  primaryclass = {gr-qc},
  note = {[arXiv:2403.10048 [gr-qc]]},
}

@article{Doneva:2019vuh,
  author = {D.~D.~Doneva and K.~V.~Staykov and S.~S.~Yazadjiev},
  title = {{Gauss-Bonnet black holes with a massive scalar field}},
  journal = {Phys. Rev. D},
  volume = {99},
  pages = {104045},
  year = {2019},
  eprint = {1903.08119},
  archiveprefix = {arXiv},
  primaryclass = {gr-qc},
  note = {[arXiv:1903.08119 [gr-qc]]},
}

@article{Macedo:2019sem,
  author = {C.~F.~B.~Macedo and J.~Sakstein and E.~Berti and L.~Gualtieri and H.~O.~Silva and T.~P.~Sotiriou},
  title = {{Self-interactions and spontaneous black hole scalarization}},
  journal = {Phys. Rev. D},
  volume = {99},
  pages = {104041},
  year = {2019},
  eprint = {1903.06784},
  archiveprefix = {arXiv},
  primaryclass = {gr-qc},
  note = {[arXiv:1903.06784 [gr-qc]]},
}

@article{Peng:2020znl,
  author = {Y.~Peng},
  title = {{Spontaneous scalarization of Gauss-Bonnet black holes surrounded by massive scalar fields}},
  journal = {Phys. Lett. B},
  volume = {807},
  pages = {135569},
  year = {2020},
  eprint = {2004.12566},
  archiveprefix = {arXiv},
  primaryclass = {gr-qc},
  note = {[arXiv:2004.12566 [gr-qc]]},
}

@article{Macedo:2020tbm,
  author = {C.~F.~B.~Macedo},
  title = {{Scalar modes, spontaneous scalarization and circular null-geodesics of black holes in scalar-Gauss-Bonnet gravity}},
  journal = {Int. J. Mod. Phys. D},
  volume = {29},
  pages = {2041006},
  year = {2020},
  eprint = {2002.12719},
  archiveprefix = {arXiv},
  primaryclass = {gr-qc},
  note = {[arXiv:2002.12719 [gr-qc]]},
}

@article{Cunha:2019dwb,
  author = {P.~V.~P.~Cunha and C.~A.~R.~Herdeiro and E.~Radu},
  title = {{Spontaneously scalarized Kerr black holes in extended scalar-tensor-Gauss-Bonnet gravity}},
  journal = {Phys. Rev. Lett.},
  volume = {123},
  pages = {011101},
  year = {2019},
  eprint = {1904.09997},
  archiveprefix = {arXiv},
  primaryclass = {gr-qc},
  note = {[arXiv:1904.09997 [gr-qc]]},
}

@article{Collodel:2019kkx,
  author = {L.~G.~Collodel and B.~Kleihaus and J.~Kunz and E.~Berti},
  title = {{Spinning and excited black holes in Einstein-scalar-Gauss-Bonnet theory}},
  journal = {Class. Quant. Grav.},
  volume = {37},
  pages = {075018},
  year = {2020},
  eprint = {1912.05382},
  archiveprefix = {arXiv},
  primaryclass = {gr-qc},
  note = {[arXiv:1912.05382 [gr-qc]]},
}

@article{Dima:2020yac,
  author = {A.~Dima and E.~Barausse and N.~Franchini and T.~P.~Sotiriou},
  title = {{Spin-induced black hole spontaneous scalarization}},
  journal = {Phys. Rev. Lett.},
  volume = {125},
  pages = {231101},
  year = {2020},
  eprint = {2006.03095},
  archiveprefix = {arXiv},
  primaryclass = {gr-qc},
  note = {[arXiv:2006.03095 [gr-qc]]},
}

@article{Hod:2020jjy,
  author = {S.~Hod},
  title = {{Onset of spontaneous scalarization in spinning Gauss-Bonnet black holes}},
  journal = {Phys. Rev. D},
  volume = {102},
  pages = {084060},
  year = {2020},
  eprint = {2006.09399},
  archiveprefix = {arXiv},
  primaryclass = {gr-qc},
  note = {[arXiv:2006.09399 [gr-qc]]},
}

@article{Doneva:2020kfv,
  author = {D.~D.~Doneva and L.~G.~Collodel and C.~J.~Kr\"uger and S.~S.~Yazadjiev},
  title = {{Spin-induced scalarization of Kerr black holes with a massive scalar field}},
  journal = {Eur. Phys. J. C},
  volume = {80},
  pages = {1205},
  year = {2020},
  eprint = {2009.03774},
  archiveprefix = {arXiv},
  primaryclass = {gr-qc},
  note = {[arXiv:2009.03774 [gr-qc]]},
}

@article{Doneva:2020nbb,
  author = {D.~D.~Doneva and L.~G.~Collodel and C.~J.~Kr\"uger and S.~S.~Yazadjiev},
  title = {{Black hole scalarization induced by the spin: 2+1 time evolution}},
  journal = {Phys. Rev. D},
  volume = {102},
  pages = {104027},
  year = {2020},
  eprint = {2008.07391},
  archiveprefix = {arXiv},
  primaryclass = {gr-qc},
  note = {[arXiv:2008.07391 [gr-qc]]},
}

@article{Herdeiro:2020wei,
  author = {C.~A.~R.~Herdeiro and E.~Radu and H.~O.~Silva and T.~P.~Sotiriou and N.~Yunes},
  title = {{Spin-induced scalarized black holes}},
  journal = {Phys. Rev. Lett.},
  volume = {126},
  pages = {011103},
  year = {2021},
  eprint = {2009.03904},
  archiveprefix = {arXiv},
  primaryclass = {gr-qc},
  note = {[arXiv:2009.03904 [gr-qc]]},
}

@article{Berti:2020kgk,
  author = {E.~Berti and L.~G.~Collodel and B.~Kleihaus and J.~Kunz},
  title = {{Spin-induced black-hole scalarization in Einstein-scalar-Gauss-Bonnet theory}},
  journal = {Phys. Rev. Lett.},
  volume = {126},
  pages = {011104},
  year = {2021},
  eprint = {2009.03905},
  archiveprefix = {arXiv},
  primaryclass = {gr-qc},
  note = {[arXiv:2009.03905 [gr-qc]]},
}

@article{Doneva:2022yqu,
  author = {D.~D.~Doneva and L.~G.~Collodel and S.~S.~Yazadjiev},
  title = {{Spontaneous nonlinear scalarization of Kerr black holes}},
  journal = {Phys. Rev. D},
  volume = {106},
  pages = {104027},
  year = {2022},
  eprint = {2208.02077},
  archiveprefix = {arXiv},
  primaryclass = {gr-qc},
  note = {[arXiv:2208.02077 [gr-qc]]},
}

@article{Lai:2023gwe,
  author = {M.~Y.~Lai and D.~C.~Zou and R.~H.~Yue and Y.~S.~Myung},
  title = {{Nonlinearly scalarized rotating black holes in Einstein-scalar-Gauss-Bonnet theory}},
  journal = {Phys. Rev. D},
  volume = {108},
  pages = {084007},
  year = {2023},
  eprint = {2304.08012},
  archiveprefix = {arXiv},
  primaryclass = {gr-qc},
  note = {[arXiv:2304.08012 [gr-qc]]},
}

@article{Lai:2022ppn,
  author = {M.~Y.~Lai and Y.~S.~Myung and R.~H.~Yue and D.~C.~Zou},
  title = {{Spin-charge induced spontaneous scalarization of Kerr-Newman black holes}},
  journal = {Phys. Rev. D},
  volume = {106},
  pages = {084043},
  year = {2022},
  eprint = {2208.11849},
  archiveprefix = {arXiv},
  primaryclass = {gr-qc},
  note = {[arXiv:2208.11849 [gr-qc]]},
}

@article{Staykov:2022uwq,
  author = {K.~V.~Staykov and D.~D.~Doneva},
  title = {{Multiscalar Gauss-Bonnet gravity: scalarized black holes beyond spontaneous scalarization}},
  journal = {Phys. Rev. D},
  volume = {106},
  pages = {104064},
  year = {2022},
  eprint = {2209.01038},
  archiveprefix = {arXiv},
  primaryclass = {gr-qc},
  note = {[arXiv:2209.01038 [gr-qc]]},
}

@article{Staykov:2024jbq,
  author = {K.~V.~Staykov and D.~D.~Doneva},
  title = {{Nonlinear black hole scalarization in multi-scalar Gauss-Bonnet gravity}},
  journal = {J. Phys. Conf. Ser.},
  volume = {2719},
  pages = {012007},
  year = {2024},
}

@article{Liu:2024bzh,
  author = {H.~S.~Liu and L.~Zhang},
  title = {{Scalarization of Taub-NUT black holes in extended scalar-tensor-Gauss-Bonnet theory}},
  journal = {JHEP},
  volume = {10},
  pages = {067},
  year = {2024},
  eprint = {2407.08208},
  archiveprefix = {arXiv},
  primaryclass = {gr-qc},
  note = {[arXiv:2407.08208 [gr-qc]]},
}

@article{Liu:2025eve,
  author = {S.~Liu and Y.~Liu and Y.~Peng and C.~Y.~Zhang},
  title = {{Non-linearly scalarized supermassive black holes}},
  journal = {Eur. Phys. J. C},
  volume = {85},
  pages = {1370},
  year = {2025},
  doi = {10.1140/epjc/s10052-025-15096-8},
  eprint = {2509.17892},
  archiveprefix = {arXiv},
  primaryclass = {gr-qc},
  note = {[arXiv:2509.17892 [gr-qc]]},
}

@article{Doneva:2021tvn,
  author = {D.~D.~Doneva and S.~S.~Yazadjiev},
  title = {{Beyond the spontaneous scalarization: New fully nonlinear mechanism for the formation of scalarized black holes and its dynamical development}},
  journal = {Phys. Rev. D},
  volume = {105},
  pages = {L041502},
  year = {2022},
  eprint = {2107.01738},
  archiveprefix = {arXiv},
  primaryclass = {gr-qc},
  note = {[arXiv:2107.01738 [gr-qc]]},
}

@article{Zhang:2023jei,
  author = {S.~J.~Zhang},
  title = {{Nonlinear instability and scalar clouds of spherical exotic compact objects in scalar-Gauss-Bonnet theory}},
  journal = {Eur. Phys. J. C},
  volume = {83},
  pages = {950},
  year = {2023},
  eprint = {2304.08092},
  archiveprefix = {arXiv},
  primaryclass = {gr-qc},
  note = {[arXiv:2304.08092 [gr-qc]]},
}

@article{Blazquez-Salcedo:2022omw,
  author = {J.~L.~Bl\'azquez-Salcedo and D.~D.~Doneva and J.~Kunz and S.~S.~Yazadjiev},
  title = {{Radial perturbations of scalar-Gauss-Bonnet black holes beyond spontaneous scalarization}},
  journal = {Phys. Rev. D},
  volume = {105},
  pages = {124005},
  year = {2022},
  eprint = {2203.00709},
  archiveprefix = {arXiv},
  primaryclass = {gr-qc},
  note = {[arXiv:2203.00709 [gr-qc]]},
}

@article{Pombo:2023lxg,
  author = {A.~M.~Pombo and D.~D.~Doneva},
  title = {{Effects of mass and self-interaction on nonlinear scalarization of scalar-Gauss-Bonnet black holes}},
  journal = {Phys. Rev. D},
  volume = {108},
  pages = {124068},
  year = {2023},
  eprint = {2310.08638},
  archiveprefix = {arXiv},
  primaryclass = {gr-qc},
  note = {[arXiv:2310.08638 [gr-qc]]},
}

@article{Zou:2024paper,
  author = {D.~C.~Zou and X.~Yang and M.~Y.~Lai and H.~Huang and B.~Liu and J.~Kunz and Y.~S.~Myung and R.~H.~Yue},
  title = {{Existence of nonlinearly scalarized black holes in Einstein-scalar-Gauss-Bonnet theory with polynomial couplings}},
  journal = {Phys. Rev. D},
  volume = {113},
  pages = {104015},
  year = {2026},
  doi = {10.1103/y3sr-wd4q},
  eprint = {2404.19521},
  archiveprefix = {arXiv},
  primaryclass = {gr-qc},
  note = {doi:10.1103/y3sr-wd4q, [arXiv:2404.19521 [gr-qc]]},
}

@article{Li:2025vcq,
    author = {Z.~Li and H.~S.~Liu and H.~L\"u},
    title = "{From wormhole to black hole with negative mass}",
    eprint = "2509.20755",
    archivePrefix = "arXiv",
    primaryClass = "gr-qc",
    doi = "10.1103/9h7h-686p",
    journal = "Phys. Rev. D",
    volume = "113",
    number = "4",
    pages = "044032",
    year = "2026",
    note = {[arXiv:2509.20755 [gr-qc]]},
}

@article{Myung:2019radialcomparison,
  author = {Y.~S.~Myung and D.~C.~Zou},
  title = {{Stability of scalarized charged black holes in the Einstein-Maxwell-Scalar theory}},
  year = {2019},
  eprint = {1904.09864},
  archivePrefix = {arXiv},
  primaryClass = {gr-qc},
  journal = {Eur. Phys. J. C},
  volume = {79},
  pages = {641},
  doi = {10.1140/epjc/s10052-019-7176-7},
  note = {[arXiv:1904.09864 [gr-qc]]},
}

@article{Kimura:2018whv,
  author = {M.~Kimura and T.~Tanaka},
  title = {{Robustness of the S-deformation method for black hole stability analysis}},
  journal = {Class. Quant. Grav.},
  volume = {35},
  pages = {195008},
  year = {2018},
  doi = {10.1088/1361-6382/aadc13},
  eprint = {1805.08625},
  archivePrefix = {arXiv},
  primaryClass = {gr-qc},
  note = {[arXiv:1805.08625 [gr-qc]]},
}

@article{Caldwell:1999ew,
  author = {R.~R.~Caldwell},
  title = {{A Phantom menace?}},
  journal = {Phys. Lett. B},
  volume = {545},
  pages = {23},
  year = {2002},
  eprint = {astro-ph/9908168},
  archivePrefix = {arXiv},
  note = {[arXiv:astro-ph/9908168]},
}

@article{Babichev:2004yx,
  author = {E.~Babichev and V.~Dokuchaev and Y.~Eroshenko},
  title = {{Black hole mass decreasing due to phantom energy accretion}},
  journal = {Phys. Rev. Lett.},
  volume = {93},
  pages = {021102},
  year = {2004},
  eprint = {gr-qc/0402089},
  archivePrefix = {arXiv},
  note = {[arXiv:gr-qc/0402089]},
}

@article{Ellis:1973yv,
  author = {H.~G.~Ellis},
  title = {{Ether flow through a drainhole - a particle model in general relativity}},
  journal = {J. Math. Phys.},
  volume = {14},
  pages = {104},
  year = {1973},
}

@article{Bronnikov:1973fh,
  author = {K.~A.~Bronnikov},
  title = {{Scalar-tensor theory and scalar charge}},
  journal = {Acta Phys. Polon. B},
  volume = {4},
  pages = {251},
  year = {1973},
}

@article{Bronnikov:2005gm,
  author = {K.~A.~Bronnikov and J.~C.~Fabris},
  title = {{Regular phantom black holes}},
  journal = {Phys. Rev. Lett.},
  volume = {96},
  pages = {251101},
  year = {2006},
  eprint = {gr-qc/0511109},
  archivePrefix = {arXiv},
  note = {[arXiv:gr-qc/0511109]},
}

@article{Clement:2009ai,
  author = {G.~Cl\'ement and J.~C.~Fabris and M.~E.~Rodrigues},
  title = {{Phantom black holes in Einstein-Maxwell-dilaton theory}},
  journal = {Phys. Rev. D},
  volume = {79},
  pages = {064021},
  year = {2009},
  eprint = {0901.4543},
  archivePrefix = {arXiv},
  primaryClass = {hep-th},
  note = {[arXiv:0901.4543 [hep-th]]},
}

@article{Gonzalez:2008wd,
  author = {J.~A.~Gonz\'alez and F.~S.~Guzm\'an and O.~Sarbach},
  title = {{Instability of wormholes supported by a ghost scalar field. I. Linear stability analysis}},
  journal = {Class. Quant. Grav.},
  volume = {26},
  pages = {015010},
  year = {2009},
  eprint = {0806.0608},
  archivePrefix = {arXiv},
  primaryClass = {gr-qc},
  note = {[arXiv:0806.0608 [gr-qc]]},
}

@article{Bronnikov:2012ch,
  author = {K.~A.~Bronnikov and R.~A.~Konoplya and A.~Zhidenko},
  title = {{Instabilities of wormholes and regular black holes supported by a phantom scalar field}},
  journal = {Phys. Rev. D},
  volume = {86},
  pages = {024028},
  year = {2012},
  eprint = {1205.2224},
  archivePrefix = {arXiv},
  primaryClass = {gr-qc},
  note = {[arXiv:1205.2224 [gr-qc]]},
}

@misc{Zou:2026staticthermo,
  author = {M.~Zhang and D.~C.~Zou and Y.~S.~Myung and B.~Liu and Y.~Long and L.~Zhang},
  title = {{Thermodynamics and phase transitions of nonlinearly scalarized black holes in Einstein-phantom scalar-Gauss-Bonnet theory}},
  year = {2026},
  eprint = {2609.39236},
  archivePrefix = {arXiv},
  primaryClass = {gr-qc},
  url = {https://arxiv.org/abs/2609.39236v3},
  note = {[arXiv:2609.39236[gr-qc]]},
}

@article{Zou:2026thermo,
  author = {D.~C.~Zou and X.~Yang and M.~Y.~Lai and H.~Huang and Y.~S.~Myung},
  title = {{Thermodynamics and phase transitions of nonlinearly scalarized black holes in Einstein-scalar-Gauss-Bonnet theory}},
  journal = {Phys. Lett. B},
  volume = {878},
  pages = {140573},
  year = {2026},
  eprint = {2604.20153},
  archiveprefix = {arXiv},
  primaryclass = {gr-qc},
  note = {[arXiv:2604.20153 [gr-qc]]},
}

@article{Herdeiro:2026sur,
  author = {C.~Herdeiro and H.~Huang and J.~Kunz and M.~Y.~Lai and E.~Radu and D.~C.~Zou},
  title = {{Phase structure of scalarized black holes in Einstein-scalar-Gauss-Bonnet gravity}},
  journal = {Phys. Rev. D},
  volume = {113},
  pages = {124046},
  year = {2026},
  doi = {10.1103/4k1l-jzvz},
  eprint = {2603.24164},
  archiveprefix = {arXiv},
  primaryclass = {gr-qc},
  note = {[arXiv:2603.24164 [gr-qc]]},
}

\endgroup

\end{document}